%% file: 00_paper.tex
\documentclass[runningheads]{llncs}
\usepackage[T1]{fontenc}
\usepackage{url}
\usepackage{graphicx}

\newcommand{\XPERCENT}{38\% }
\newcommand{\YPERCENT}{96\% }
\newcommand{\ZPERCENT}{96\% }
\begin{document}
\title{Is ChatGPT 3 safe for students?}
%
%
\author{Julia Kotovich\inst{1} \and Manuel Oriol\inst{1}\orcidID{0000-0003-4069-7626}}
\authorrunning{J. Kotovich and M. Oriol}
%
\institute{Constructor Institute, Schaffhausen, Switzerland,
\email{julia.kotovich@constructor.org, mo@constructor.org}}
\maketitle              
\begin{abstract}
ChatGPT3 is a chat engine that fulfils the promises of an AI-based chat engine: users can ask a question (prompt) and it answers in a reasonable manner. 
The coding-related skills of ChatGPT are especially impressive: informal testing shows that it is difficult to find simple questions that ChatGPT3 does not know how to answer properly. 
Some students are certainly already using it to answer programming assignments. 

This article studies whether it is safe for students to use ChatGPT3 to answer coding assignments (``safe'' means that they will not be caught for plagiarism if they use it).
The main result is that it is generally not safe for students to use ChatGPT3. 
We evaluated the safety of code generated with ChatGPT3, by performing a search with a Codequiry, a plagiarism detection tool, and searching plagiarized code in Google (only considering the first page of results).
In \XPERCENT of the cases, Codequiry finds a piece of code that is partially copied by the answer of ChatGPT3. 
In \YPERCENT of the cases, the Google search finds a piece of code very similar to the generated code.
Overall, it is not safe for students to use ChatGPT3 in \ZPERCENT of the cases.  

\keywords{ChatGPT \and education \and programming.}
\end{abstract}
%
%
%
%




\input{01_introduction}

\input{02_experiment}
\input{03_results}
\input{04_threat_to_validity}

\input{05_related_work}

\input{06_conclusions}

\bibliography{biblio}
\bibliographystyle{splncs04}
\end{document}

%% file: 01_introduction.tex
\section{Introduction}\label{sec:intro}
In the past few months, ChatGPT3\footnote{\url{https://openai.com/blog/chatgpt}} has been at the heart of many discussions between academics because of its potential to change what educators can ask students to code.
It looks like many simple coding tasks can be automatically performed using ChatGPT3.
Simple assignments can then be solved in a very short amount of time without understanding the generated code, and with learning only one skill: asking the right questions to the chat engine.
This article investigates whether students can safely write programs using ChatGPT3 for assignments that forbid it.

Academics have predicted that ChatGPT will change the way software engineers code~\cite{10.1145/3555367}, or even kill programming altogether~\cite{10.1145/3570220}.
To adapt to the new paradigm, educators will change the way they teach in the future.
Adapting the education programs and expectations will however take time and some students already started to use the new technology. 
This poses two main challenges: (1) Will ChatGPT actually produce the right answer? (2) Will educators be able to detect its use?

This article evaluates these two challenges by using ChatGPT3 on programming tasks that consist in coding standard data structures and standard sorting algorithms. 
Such algorithms are very well documented and generally available online.
Students who want to cheat can already use resources online, but they generally need to adapt them to fit the programming language or the data format. 
This article's main experiment consists in asking ChatGPT3 to code algorithms from BigOCheatSheet,\footnote{\url{https://www.bigocheatsheet.com}} check whether a standard plagiarism tool (Codequiry~\cite{CodeQuiry}) detects it, and then check whether the first page of a simple Google search returns results that can be referenced to show plagiarism.
It is then considered \emph{safe} for students to use ChatGPT3 if our study does not find any data that shows plagiarism.

The main results of this study are that ChatGPT produced the correct answers 100\% of the time for that basic standard requests in computer science.
However, it is generally not safe for students to use ChatGPT3 to generate these simple algorithms. 
In \XPERCENT of the cases, Codequiry finds a piece of code that is partially copied by the answer of ChatGPT3. 
In \YPERCENT of the cases anyway, the Google search finds a piece of code that is very similar to the generated code.
Overall, it is not safe for students to use ChatGPT3 in \ZPERCENT of the cases. 

Section~\ref{sec:high} describes the experiment in more details.
Section~\ref{sec:xp} presents the main findings and their implications.
Section~\ref{sec:threat} explains the threats to validity.
Section~\ref{sec:rw} analyzes related work.
Section~\ref{sec:conc} concludes this study.

%% file: 02_experiment.tex
\section{Experiment}\label{sec:high}
The experiment consists in simulating an assignment made by a lecturer requiring students to code a standard algorithm in Python.

\textbf{Why Python?} Python is the most popular general-purpose programming language on StackOverflow.\footnote{\url{https://insights.stackoverflow.com/survey/2021#technology-most-popular-technologies}}

\textbf{How did we select the algorithms?} 
Algorithms selected for the test correspond to the most used data structure and sorting algorithms presented on the BigOCheatSheet (the first result on a Google search that describes "algorithms complexity" with a complete list of algorithms, January 2023).

This results in a list of 13 sorting algorithms and 14 data structures:
\begin{enumerate}
    \item{Quicksort}
    \item{Mergesort}
    \item{Timsort}
    \item{Heapsort}
    \item{Bubble Sort}
    \item{Insertion Sort}
    \item{Selection Sort}
    \item{Tree Sort}
    \item{Shell Sort}
    \item{Bucket Sort}
    \item{Radix Sort}
    \item{Counting Sort}
    \item{Cubesort}
    \item{Array} (removed from the evaluation as it is a base type in Python)
    \item{Stack}
    \item{Queue}
    \item{Singly-Linked List}
    \item{Doubly-Linked List}
    \item{Skip List}
    \item{Hash Table}
    \item{Binary Search Tree}
    \item{Cartesian Tree}
    \item{B-Tree}
    \item{Red-Black Tree}
    \item{Splay Tree}
    \item{AVL Tree}
    \item{KD Tree}
\end{enumerate}

For each algorithm and data structure, (except the Array type, which is a base type in Python), we requested ChatGPT3 to create the implementation in Python using common prompt\emph{ "write a Python code for X"} (for example\emph{ "write a python code for Bubble Sort"}). All the results are then stored in a GitHub repository.\footnote{\url{https://github.com/Julia-Kotovich/ChatGPT_Python_code}}
Additionally, for each piece of code, ChatGPT provides a code snippet and a short comment such as: \emph{"This code sorts an input array 'arr' using the bubble sort algorithm. The algorithm compares each pair of adjacent elements and swaps them in they are if the wrong order. This process is repeated until the array is sorted in ascending order."}

The resulting generated code is then uploaded to CodeQuiry~\cite{CodeQuiry}. 
CodeQuiry's "Web Check tool (Checking Engine - Web Plagiarism and Group Similarity)" is a testing tool for plagiarism that compares code to over 100 million sources of code from major public and private repositories, as well as over 2 billion pieces of code from the web. 
The results show similarities and highlighted matches to external sources.
CodeQuiry \footnote{\url{https://codequiry.com/code-plagiarism}} returns the list of sources with links for a specific piece of code and the percentage of matches.

We then perform a search on the Internet using Google search and only look at the first page of results for duplicated pieces of code.
If the code is significantly duplicated (if more than 50\% of lines of code have a match or just variable names were changed), we then consider that the piece of code generated by ChatGPT3 is not safe for use by students as an answer in the assignments.



%% file: 03_results.tex
\section{Results}\label{sec:xp}

\begin{table}[h]
\caption{Codequiry results on the whole repository}\label{Codequiry1}
\begin{center}
\begin{tabular}{|l|l|l|l|}
\hline
                & Sources Indexed & Total Matches & Parseable Lines of Code \\ \hline
Algorithms      & 25,664,586,829  & 32            & 261                     \\ \hline
Data Structures & 25,590,145,664  & 45            & 632                     \\ \hline
\end{tabular}
\end{center}
\end{table}

In all cases, the code generated by ChatGPT is valid and could be used as such without any modifications.

All results of the code generated are stored in a GitHub repository and then uploaded to the CodeQuiry platform for the tests.
CodeQuiry found 32 matches (see Table~\ref{Codequiry1}) for the algorithm folder which contains 13 files there and 45 matches for the data structure folder which contains 13 files. 

In total, Codequiry finds 77 matches in the 26 files. The percentage of matching varies from 8\% to 96\% and the average is only 38\%.

CodeQuiry found two main sources for the code are StackOverflow and GitHub. Details of the results are available in Table~\ref{mainresults}.

\begin{table}[htbp]
\caption{Complete results of the evaluation. In 38\% of the cases Codequiry finds more than 50\% of the code is copied. In 96\% of the cases, a simple Google search finds a source for at least 50\% of the code.}\label{mainresults}
\begin{center}
\begin{tabular}{|l|l|l|l|l|l|l|}
\hline
\textbf{Name}      & \textbf{Codequiry\%} & \textbf{Source link}  & \textbf{Google link} & \textbf{Safe?} & \textbf{Correct?} \\ \hline
Quicksort & 69 & GitHub & rb.gy/9myp & no & yes \\ \hline
Mergesort & 96 & StackOverflow & rb.gy/qcy3 & no & yes \\ \hline
Timsort & 0 & no matches & rb.gy/bkdq & no & yes \\ \hline
Heapsort & 96 & StackOverflow & rb.gy/un4e & no & yes \\ \hline
Bubble Sort & 70 & StackOverflow & rb.gy/elp9 & no& yes \\ \hline
Insertion Sort & 75 & StackOverflow & rb.gy/wsq5 & no & yes \\ \hline
Selection Sort & 79 & GitHub & rb.gy/emlj & no & yes \\ \hline
Tree Sort & 0 & no matches & rb.gy/e1sc & no & yes \\ \hline
Shell Sort & 74 & StackOverflow & rb.gy/zhjp & no & yes \\ \hline
Bucket Sort & 0 & no matches & rb.gy/kuka & no & yes \\ \hline
Radix Sort & 78 & GitHub & rb.gy/gudz & no & yes \\ \hline
Counting Sort & 0 & no matches & rb.gy/cnnb & no & yes \\ \hline
Cubesort & 0 & no matches & rb.gy/2m5i & no & yes \\ \hline
Stack & 0 & no matches & rb.gy/ckpc & no & yes \\ \hline
Queue & 0 & no matches & rb.gy/zkft & no & yes \\ \hline
Singly-Linked List & 0 & no matches & rb.gy/o3iy & no & yes \\ \hline
Doubly-Linked List & 14 & GitHub & rb.gy/30c7 & no & yes \\ \hline
Skip List & 0 & no matches & rb.gy/dfvc & no & yes \\ \hline
Hash Table & 0 & no matches & no matches   & yes & yes \\ \hline
Binary Search Tree & 10 & GitHub & rb.gy/tgcs & no & yes \\ \hline
Cartesian Tree & 0 & no matches & rb.gy/2af2 & no & yes \\ \hline
B-Tree & 0 & no matches & rb.gy/oed8 & no & yes \\ \hline
Red-Black Tree & 22 & GitHub & rb.gy/i9vm & no & yes \\ \hline
Splay Tree & 8 & StackOverflow & rb.gy/apoi & no & yes \\ \hline
AVL Tree & 58 & GitHub & rb.gy/ibu0 & no & yes \\ \hline
KD Tree & 12  & StackOverflow & rb.gy/upak & no & yes \\ \hline
\end{tabular}
\end{center}
\end{table}

According to the results of a Google search, the most common sources that Google finds code similar to the generated by ChatGPT are these popular websites: StackOverflow, GitHub, GeeksforGeeks,\footnote{\url{https://www.geeksforgeeks.org/}} Programiz,\footnote{\url{https://www.programiz.com/}} and freeCodeCamp.\footnote{\url{https://www.freecodecamp.org/}}

In only \XPERCENT of the cases Codequiry finds the similarity. 
The results of a manual Google search revealed that in \ZPERCENT of the cases, code or portions of code very similar to what was generated using ChatGPT could be found on the first page of Google results.
This results in an overall \ZPERCENT of the cases being unsafe.




%% file: 04_threat_to_validity.tex
\section{Limitations and Threats to Validity}\label{sec:threat}
There are mainly four threats to the validity of this study: (1) the algorithms used to test ChatGPT3 are not representative of "code in general", (2) generated code is always the same for the same question, (3) the methodology for considering safety is not evaluating the right actions, and (4) these results are only limited to ChatGPT3.
The following paragraphs evaluate each one of these threats separately.

\paragraph{Threat 1: Algorithms are not representative.} 
The question is whether students can use ChatGPT to code a correct solution. 
The answer is that, in most cases, they can.
This seems true for simple algorithms, but what happens for smarter programming questions? 
What about questions that require to combine several aspects of such algorithms?
Does ChatGPT3 still generate the correct code?
Is it still safe?
Our guess is that the more complicated the problem, the more likely it is that ChatGPT3 does not produce the right result though it might be more difficult to spot plagiarism.
This should, however, be checked in further studies.

\paragraph{Threat 2: Generated code is always the same.} 
When one asks questions, ChatGPT3 answers. If the same question is asked again ChatGPT3 might return another answer!
Our preliminary data show that these answers have strong similarities (around 60\% for the few cases that we evaluated).
It is however possible that these might diverge significantly and that our conclusions are erroneous because of that.
Again this would deserve a further study.

\paragraph{Threat 3: Actions are wrongly chosen to evaluate the safety of the approach.}
In many universities, plagiarism tools are simply not used, and we cannot imagine a coding instructor checking all projects one by one using Google, and hunting for references. 
This means that the study has conservative conclusions.
Since we are considering safety, it seems appropriate to be conservative, especially considering that plagiarism tools might improve and catch even more issues in the future. 


\paragraph{Threat 4: The study only applies to ChatGPT3.}
It is clear that this article only considered ChatGPT3 as a target because it seemed to be the best-adapted tool when we started the study. 
Since then, ChatGPT4 appeared, and we did not study other tools.
We believe that, because all these tools are learning from the same sources, we would see similar results with other tools. 
We however have no evidence of that fact.

%% file: 05_related_work.tex
\section{Related Work}\label{sec:rw}
Bots are becoming more and more available to software engineers willing to improve their productivity~\cite{8823643,mapping}.
For example, Carr \textit{et al.}~\cite{DBLP:journals/tse/CarrLP17} created a bot that inserts automatically proven contracts in source code, Tian \textit{et al.}~\cite{8115628} made a chatbot that answers questions about APIs, Bradley \textit{et al.}~\cite{10.1145/3180155.3180238} made a development assistant able to understand commands for Git and GitHub tasks.

For automated bots generating code, most articles tend to focus on making it as close to what a programmer could have generated. For example, generating automatically patches with explanations~\cite{8823632,10.1145/3183519.3183540} or make refactorings indistinguishable from what a human could have generated~\cite{8823629}.

Plagiarism of source code is a widely studied field~\cite{7522248} that focused mostly on detecting plagiarism among a group of students who received the same question.

Internet plagiarism commercial detection tools like Codequiry~\cite{CodeQuiry}, copyleaks~\cite{copyleaks}, or Turnitin~\cite{turnitin} are already promising to detect AI-generated content. 
In our preliminary evaluation, they are not yet accurate enough to exhibit a good accuracy. 
It is however clear that these tools will be able to detect the code and content generated by the current generation of AI engines.
To our knowledge, we are the first ones to present a study showing that such content can be detected.


%% file: 06_conclusions.tex
\section{Conclusions and Future Work}\label{sec:conc}
Some tools suddenly open possibilities that we thought would never be reality.
ChatGPT3 is one of these tools.
It suddenly sparked a very strong interest and captured the imagination of many.
Is it interesting? Yes! 
It seems to return only valid results, which could be expected for simple cases, but is it \textit{safe} to use for programming assignments? No!

If students use ChatGPT3 for simple assignments, they have very high chances to be penalised for plagiarism.
Even if the plagiarism tool we used only found plagiarism in \XPERCENT  of the cases, a simple Google search finds plagiarism in \ZPERCENT of the cases.
It is likely that future versions of the Internet plagiarism finding tools improve and catch these cases better.
This might lead to retroactive invalidation of results (similarly to drug tests for athletes).

It is possible that other tools than ChatGPT3 create better results, but if everyone uses them, it is also possible that results coming from different students look very much alike and then are identified as plagiarism. 
Future work will focus on confirming our results using other tools to generate code and more complex requests to generate code.

%% file: 00_paper.bbl
\begin{thebibliography}{10}
\providecommand{\url}[1]{\texttt{#1}}
\providecommand{\urlprefix}{URL }
\providecommand{\doi}[1]{https://doi.org/#1}

\bibitem{10.1145/3180155.3180238}
Bradley, N.C., Fritz, T., Holmes, R.: Context-aware conversational developer
  assistants. In: Proceedings of the 40th International Conference on Software
  Engineering. p. 993–1003. ICSE '18, Association for Computing Machinery,
  New York, NY, USA (2018). \doi{10.1145/3180155.3180238},
  \url{https://doi.org/10.1145/3180155.3180238}

\bibitem{DBLP:journals/tse/CarrLP17}
Carr, S.A., Logozzo, F., Payer, M.: Automatic contract insertion with ccbot.
  {IEEE} Trans. Software Eng.  \textbf{43}(8),  701--714 (2017).
  \doi{10.1109/TSE.2016.2625248},
  \url{https://doi.org/10.1109/TSE.2016.2625248}

\bibitem{CodeQuiry}
CodeQuiry, L.: Codequiry. \url{https://codequiry.com} (2023 (accessed February,
  2023))

\bibitem{copyleaks}
Copyleaks, I.: Copyleaks. \url{https://www.copyleaks.com/} (2023 (accessed
  March, 2023))

\bibitem{8823643}
Erlenhov, L., Gomes~de Oliveira~Neto, F., Scandariato, R., Leitner, P.: Current
  and future bots in software development. In: 2019 IEEE/ACM 1st International
  Workshop on Bots in Software Engineering (BotSE). pp. 7--11 (2019).
  \doi{10.1109/BotSE.2019.00009}

\bibitem{8823632}
Monperrus, M.: Explainable software bot contributions: Case study of automated
  bug fixes. In: 2019 IEEE/ACM 1st International Workshop on Bots in Software
  Engineering (BotSE). pp. 12--15. IEEE Computer Society, Los Alamitos, CA, USA
  (may 2019). \doi{10.1109/BotSE.2019.00010},
  \url{https://doi.ieeecomputersociety.org/10.1109/BotSE.2019.00010}

\bibitem{7522248}
Novak, M.: Review of source-code plagiarism detection in academia. In: 2016
  39th International Convention on Information and Communication Technology,
  Electronics and Microelectronics (MIPRO). pp. 796--801 (2016).
  \doi{10.1109/MIPRO.2016.7522248}

\bibitem{mapping}
Santhanam, S., Hecking, T., Schreiber, A., Wagner, S.: Bots in software
  engineering: a systematic mapping study. PeerJ Comput Science  \textbf{8}
  (2022)

\bibitem{8115628}
Tian, Y., Thung, F., Sharma, A., Lo, D.: Apibot: Question answering bot for api
  documentation. In: 2017 32nd IEEE/ACM International Conference on Automated
  Software Engineering (ASE). pp. 153--158 (2017).
  \doi{10.1109/ASE.2017.8115628}

\bibitem{turnitin}
Turnitin, L.: Turnitin. \url{https://www.turnitin.com/} (2023 (accessed March,
  2023))

\bibitem{10.1145/3183519.3183540}
Urli, S., Yu, Z., Seinturier, L., Monperrus, M.: How to design a program repair
  bot? insights from the repairnator project. In: Proceedings of the 40th
  International Conference on Software Engineering: Software Engineering in
  Practice. p. 95–104. ICSE-SEIP '18, Association for Computing Machinery,
  New York, NY, USA (2018). \doi{10.1145/3183519.3183540},
  \url{https://doi.org/10.1145/3183519.3183540}

\bibitem{10.1145/3570220}
Welsh, M.: The end of programming. Commun. ACM  \textbf{66}(1),  34–35 (dec
  2022). \doi{10.1145/3570220}, \url{https://doi.org/10.1145/3570220}

\bibitem{8823629}
Wyrich, M., Bogner, J.: Towards an autonomous bot for automatic source code
  refactoring. In: 2019 IEEE/ACM 1st International Workshop on Bots in Software
  Engineering (BotSE). pp. 24--28 (2019). \doi{10.1109/BotSE.2019.00015}

\bibitem{10.1145/3555367}
Yellin, D.M.: The premature obituary of programming. Commun. ACM
  \textbf{66}(2),  41–44 (jan 2023). \doi{10.1145/3555367},
  \url{https://doi.org/10.1145/3555367}

\end{thebibliography}
